\documentclass[twocolumn,showpacs,preprintnumbers,amsmath,amssymb]{revtex4}
\usepackage{amsmath,amssymb,graphics,epsfig,subfigure}
\usepackage{color}

\begin{document}

\thispagestyle{empty}

\begin{center}

\title{Novel dual relation and constant in Hawking-Page phase transitions}

\date{\today}
\author{Shao-Wen Wei \footnote{weishw@lzu.edu.cn, corresponding author}, Yu-Xiao Liu \footnote{liuyx@lzu.edu.cn}}
 \affiliation{Institute of Theoretical Physics $\&$ Research Center of Gravitation, Lanzhou University, Lanzhou 730000, People's Republic of China}
\author{Robert B. Mann\footnote{rbmann@uwaterloo.ca}}
\affiliation{Dept. of Physics \& Astronomy, University of Waterloo, Waterloo, Ont. Canada N2L 3G1}

\begin{abstract}
Universal relations and constants have important applications in understanding a physical theory. In this article, we explore this issue for  Hawking-Page phase transitions in Schwarzschild anti-de Sitter black holes. We find a novel exact dual relation between the minimum temperature of the ($d$+1)-dimensional black hole and the Hawking-Page phase transition temperature in $d$ dimensions, reminiscent of the holographic principle. Furthermore, we find that the normalized Ruppeiner scalar curvature is a universal constant at the Hawking-Page  transition point. Since the Ruppeiner curvature can be treated as an indicator of the intensity of the interactions amongst  black hole microstructures, we conjecture that this universal constant denotes an interaction threshold, beyond which a virtual black hole becomes a real one. This new dual relation and  universal constant are fundamental in understanding  Hawking-Page phase transitions, and might have new important applications in the black hole physics in the near future.
\end{abstract}


\maketitle
\end{center}

{\it Introduction}---Phase transitions are indicative of the competition between various internal interactions within a system and play a crucial role in contributing to our understanding of its macroscopic and microscopic properties.  This becomes particularly pertinent when universal relations and constants emerge from such investigations, the most celebrated of which are critical exponents in mean field theory that are believed to depend only on general features of a system but not on its particular details.

In black hole physics the most well-known phase transition is that observed by Hawking and Page \cite{Hawking}, which is   between a thermal radiation phase and a stable large  black hole phase in an anti-de Sitter (AdS) spacetime. This Hawking-Page (HP) phase transition  was explained as a confinement/deconfinement phase transition in gauge theory by Witten \cite{Witten2} in the AdS/CFT correspondence, which is a holographic duality between a ($d$+1)-dimensional quantum gravity and a $d$-dimensional quantum field theory \cite{Maldacena,Gubser,Witten}. It can also be understood as a solid/liquid phase transition in the context of black hole chemistry \cite{Kubiznak:2014zwa}, where the cosmological constant $\Lambda$ can regard as the system's pressure \cite{Kastor}:
\begin{equation}
 P=-\frac{\Lambda}{8\pi G_{d}}=\frac{(d-1)(d-2)}{16\pi l^2},\label{pree}
\end{equation}
where $l$ is the AdS radius. We take the $d$ dimensional gravitational constant $G_{d}$=1 for simplicity. If the black hole is charged, a complete analogy holds between the phases of a van der Waals fluid and the black hole~\cite{Mann}.

We demonstrate here some universal relations and constants associated with the HP phase transition. We find for the first time a duality relation between the minimum temperature of a Schwarzschild-AdS black hole in $(d+1)$ dimensions and the HP phase transition temperature in $d$ dimensions. This provides us with a new interpretation of the metastable large black hole phase, that we expect is indicative of some new duality relation.  We also find that the Ruppeiner scalar curvature at the HP  transition point is a  universal constant, independent of any parameters other than the spacetime dimension. Since the Ruppeiner  scalar curvature is indicative of the microscopic interactions of the constituents of a system, we conjecture that there are universal properties underlying black hole microstructures that are  key to the formation of  black holes.  We expect that our perspective on this phase transition will provide us with a foundation for  understanding the special properties of other black holes in  AdS spacetime.

\begin{figure}
\includegraphics[width=7cm]{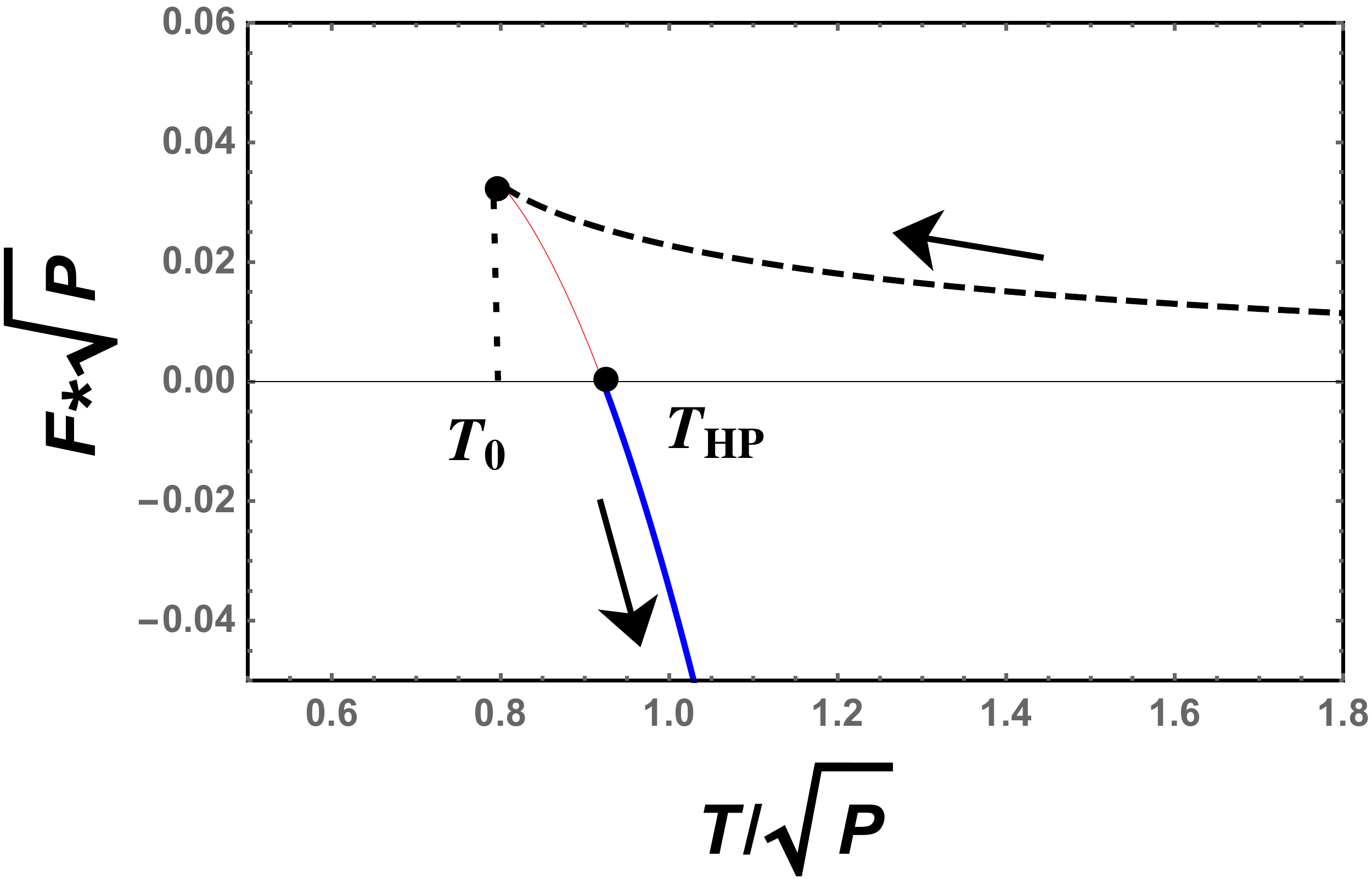}
\caption{Free energy of the four-dimensional Schwarzschild-AdS black hole. The arrows indicate  increasing  black hole horizon radius, and $T_{\text{HP}}$ and $T_0$ denote the HP phase transition  temperature and minimum temperature. The black dashed curve is the unstable small black hole branch, whereas  the metastable and stable large black hole branches are described by the red thin solid and blue solid curves, respectively.}\label{ppss}
\end{figure}

We depict the HP phase transition and its properties in Fig. \ref{ppss}. It can be seen that there are two branches corresponding to small and large black hole phases. The small black hole has negative heat capacity, and thus it is thermodynamically unstable, whereas the large black hole has positive specific heat and is stable. There exists a minimum temperature $T_0 = \sqrt{2P/\pi}$ below which no black hole can exist. In this parameter range the spacetime has only one phase, the thermal radiation phase, characterized by vanishing free energy. As the temperature of the system increases from zero, it becomes possible to form a large black hole for $T>T_0$, whose free energy is larger than that of the thermal radiation.  As such, this black hole is metastable.  Further increasing the temperature leads to a situation in which both the radiation and black hole have vanishing free energy, which is where the HP transition takes place, at  $T_{\text{HP}}$=$\sqrt{8P/3\pi}$. Above
this temperature it is thermodynamically favourable for the radiation to collapse into a   large black hole, which is the most stable phase.

The metastable large black hole branch (red thin curve in Fig. \ref{ppss}) that exists for $T_0<T<T_{\text{HP}}$ is often neglected.  However for continuous decreasing temperature it is possible for a large black hole to  pass through the HP point and for the metastable phase to emerge just like a supercooled liquid phase of water
below its freezing point. We will show that this branch  has a new and interesting interpretation.

\begin{figure}
\includegraphics[width=7cm]{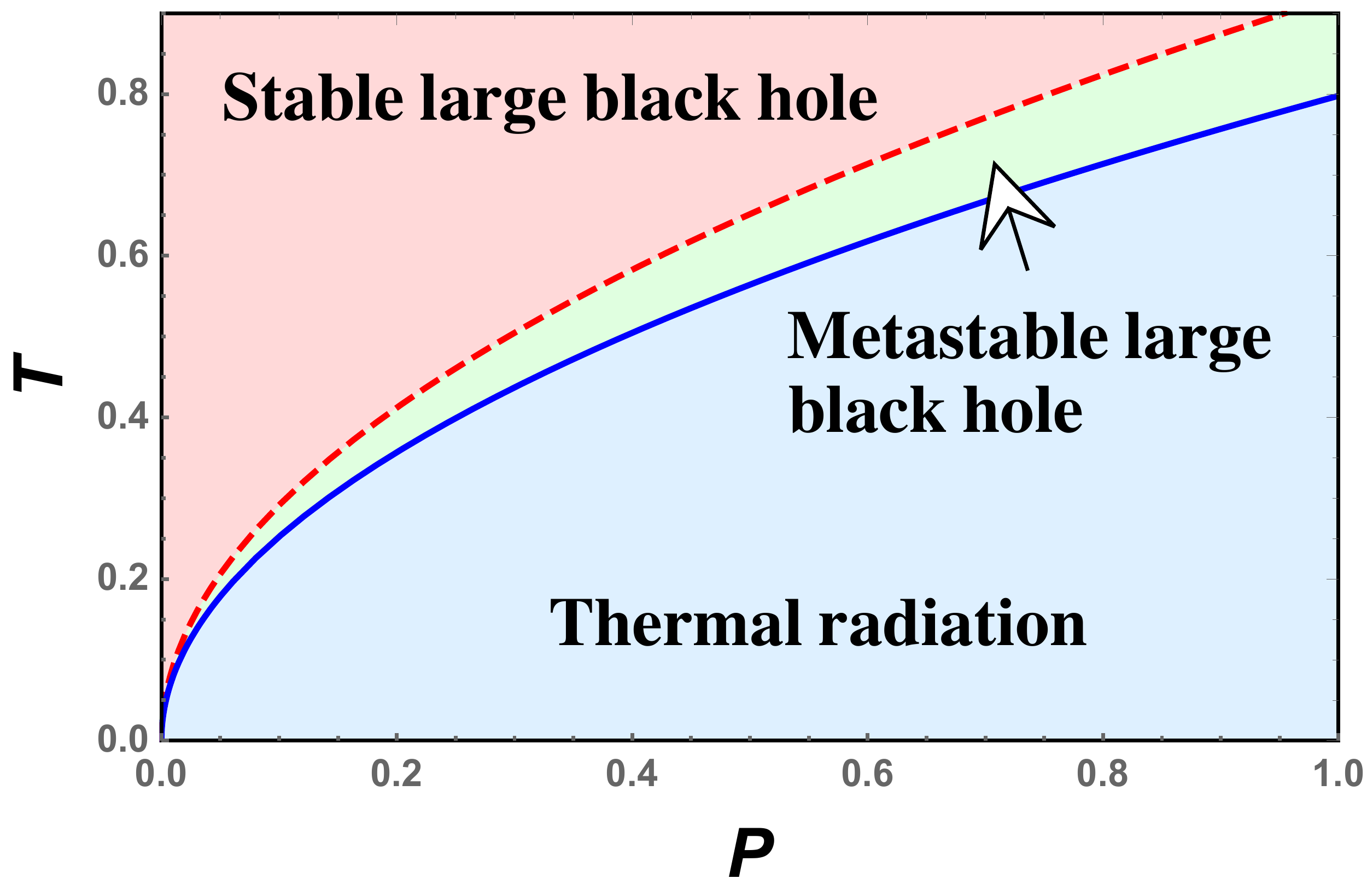}
\caption{Phase diagram for a four-dimensional Schwarzschild-AdS black hole. The blue solid and red dashed curves respectively correspond to the black hole minimum temperature and the HP phase transition temperature. Both  curves extend to infinity. }\label{pp2}
\end{figure}

{\it Novel duality in the HP  transition}--- In $d$ dimensions, the Hawking temperature of a Schwarzschild-AdS black hole is
\begin{equation}
 T=\frac{4 P r_h}{d-2}+\frac{d-3}{4 \pi  r_h},\label{tempar}
\end{equation}
with $r_h$ the radius of the black hole horizon. The pressure $P$ is defined in Eq. (\ref{pree}). The temperature is obviously dependent on the dimension of the spacetime. The free energy  has the following form \cite{Gunasekaran:2012dq,Belhaj}:
\begin{eqnarray}
 F=\frac{\omega_{d-2}  r_h^{d-3} \left((d-3) d-16 \pi
   P r_h^2+2\right)}{16 \pi  (d-2) (d-1)}, \label{free}
\end{eqnarray}
where  $\omega_{d-2}=2\pi^{(d-1)/2}/\Gamma[(d-1)/2]$ is the volume of the unit $(d-2)$-sphere. Combining (\ref{tempar}) and (\ref{free}), we can obtain the black hole minimum temperature and HP phase transition temperature
\begin{eqnarray}
 T_0&=&\sqrt{\frac{4 (d-3)}{(d-2)\pi}}\times\sqrt{P},\label{tpdh0}\\
 T_{\text{HP}}&=&\sqrt{\frac{4(d-2)}{(d-1)\pi}}\times\sqrt{P}.\label{tpdh}
\end{eqnarray}
These two temperatures share similar dependence on the pressure, $T\sim\sqrt{P}$ with different $d$-dependent coefficients. We illustrate  the complete phase diagram in Fig.~\ref{pp2} for $d=4$; the same qualitative picture holds for any $d > 4$. Indeed it is easy to see that
\begin{eqnarray}
 T_0(d)&=&\sqrt{\frac{2(d-3)}{d-2}}T_0(4),\\
 T_{\text{HP}}(d)&=&\sqrt{\frac{3(d-2)}{2(d-1)}}T_{\text{HP}}(4) \label{tpd2}
\end{eqnarray}
allowing these temperatures to be computed in any dimension from knowledge of the $d=4$ case; note that $T_{\text{HP}}/T_0$ is a decreasing function of $d$ for $d>3$. Three system phases--thermal radiation, a stable large black hole, and a metastable large black hole--are present. The minimum temperature $T_0$  and the HP  transition temperature $T_{\text{HP}}$ are respectively indicated  by the blue solid and red dashed curves. These two curves extend to infinity  \cite{Kubiznak:2014zwa}, consistent with   solid/supercooled-liquid/liquid phase behaviour at any temperature.

Simple inspection of \eqref{tpdh0} and \eqref{tpdh} indicates an underlying interesting relation
\begin{equation}
 T_{\text{HP}}(d)=T_0(d+1) \label{dual}
\end{equation}
stating that the $d$-dimensional HP phase transition temperature exactly equals to the black hole minimum temperature in one larger dimension.  This relation is also independent of the pressure $P$ and it is reminiscent of the AdS/CFT correspondence. If $T_0$ is the temperature of a physical quantity in the bulk, then $T_{\text{HP}}$ can be treated as the temperature of the dual physical quantity on the boundary. We conjecture that this universal relation will have important holographic applications and that some holographic properties should originate from it. On thing worths to emphasize is that different from the AdS/CFT correspondence,  both sides of (\ref{dual}) are in gravitational theories. Furthermore, since $T_0$ is the minimum temperature of a black hole, we can treat this as the ground  state of the AdS black hole \footnote{Note that this is not the ground state of the system if thermal radiation is taken into account.}, with the black hole at $T_{\text{HP}}$ as  one of its excited states. Thus Eq. (\ref{dual}) can also be interpreted as the duality between the ground state and an excited state of a physical system in two successive dimensions.

\begin{figure}
\includegraphics[width=6cm]{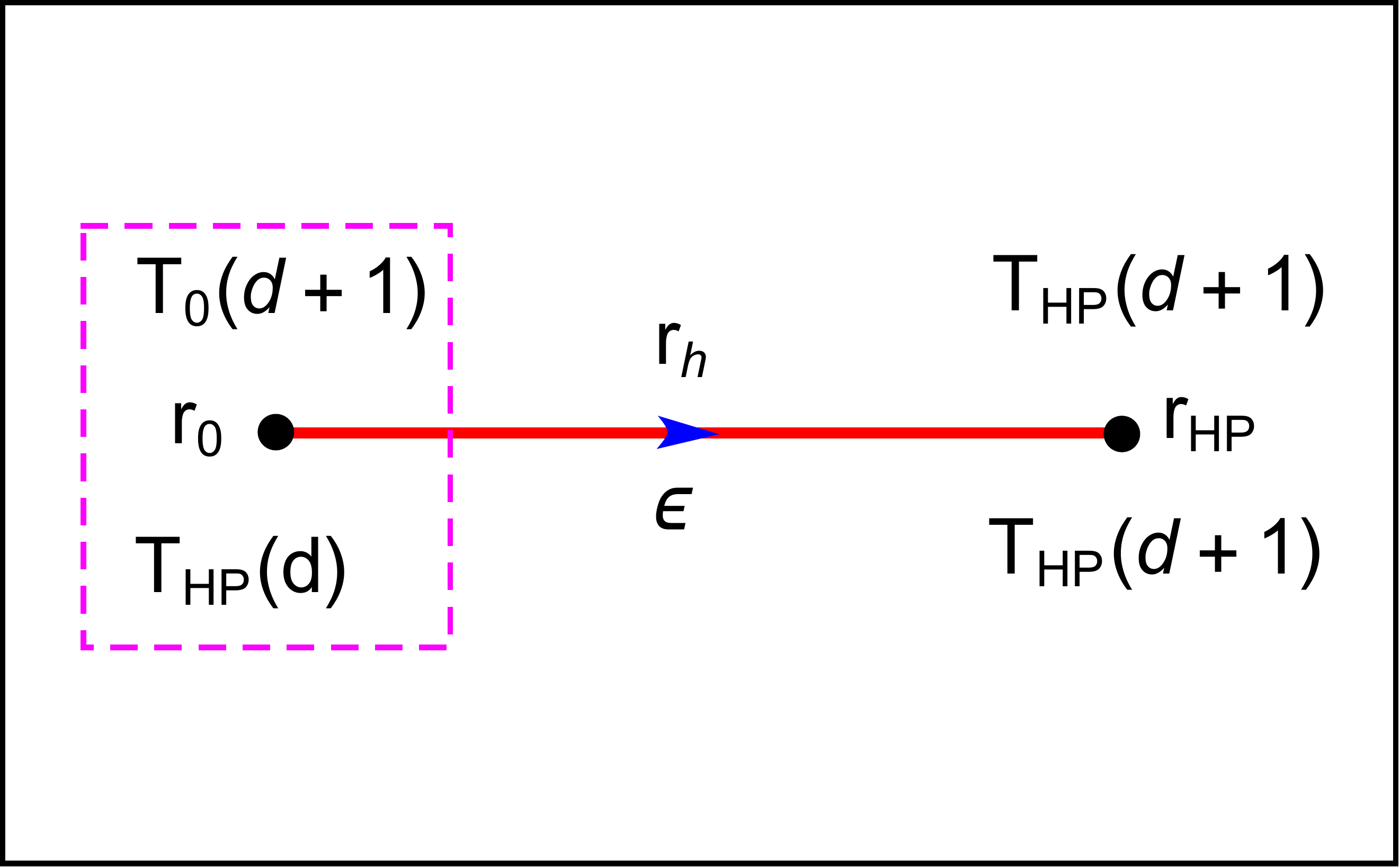}
\caption{Schematic depiction of the dual relation. The red horizontal line denotes the metastable black hole branch, with $r_0$ and $r_{\text{HP}}$ the horizon radii of the black holes at the minimum temperature and HP phase transition point respectively. The dotted box on the left is realized by the new dual relation (\ref{dual}).}\label{pd3}
\end{figure}

We sketch this dual relation in  Fig.~\ref{pd3}. The metastable large black hole branch is the horizontal line  and from  left to   right  the   horizon radius increases from $r_0$ to $r_{\text{HP}}$. Correspondingly the temperature increases from $T_0$($d$+1) to $T_{\text{HP}}$($d$+1) (see above the horizontal line).  According to the dual relation (\ref{dual}), we have $T_0(d+1)= T_{\text{HP}}(d)$. By writing
\begin{equation}
 \epsilon=\frac{r_h-r_0}{r_{\text{HP}}-r_0}\quad\in(0, 1),
\end{equation}
we see  that as $\epsilon$ varies from $0$ to $1$ we observe a new dual relation
\begin{equation}
 T_{\text{HP}}(d) \rightarrow T_{\text{HP}}(d+1),
\end{equation}
see below the horizontal line. A  similar process can also be applied to the minimum temperature of the black hole. Put succinctly,  we can say that the metastable black hole branch indeed has a physical meaning insofar as it  realizes the new dual relation $T_{\text{HP}}(d) \rightarrow T_{\text{HP}}(d+1)$ parametrized by $\epsilon$.

{\it Universal constant for black hole microstructure.}--- Understanding black hole microstructure, while a challenge in the absence of a quantum theory of gravity,
 has a wide range of applications, including for example the formations of  black holes from gravitational collapse.  Making use of
Ruppeiner geometry \cite{Ruppeiner}, constructed from thermodynamic fluctuation theory, negative or positive Ruppeiner  scalar curvature
respectively indicates (empirically) attractive or repulsive interaction among a system's microstructures.  This approach was taken for the charged AdS black hole \cite{Wei,Wei2}, which was  shown to exhibit repulsive interactions that dominate in a certain parameter range, a phenomenon that has no counterpart in a
van der Waals fluid \cite{Wei2,Wei3}, for which all microstructure interactions are attractive.

The Schwarzschild-AdS black hole is the simplest such black hole. As such we expect its  microstructures  to be fundamental, having characteristic properties
that should be shared  amongst all  AdS black holes. The line element of its thermodynamic geometry reads
\begin{equation}\label{Rupmet}
 dl^2=-\frac{1}{T}\left(\frac{\partial^2 F}{\partial T^2}\right)_{V} dT^2+\frac{1}{T}\left(\frac{\partial^2 F}{\partial V^2}\right)_{T} dV^2,
\end{equation}
where the thermodynamic volume is $V=\omega_{d-2}r_{h}^{d-1}/(d-1)$. Note that in the original Ruppeiner geometry, the black hole mass and volume, both extensive quantities, are chosen as the fluctuation coordinates, and then transformed   to   temperature and volume. Using \eqref{Rupmet} it is straightforward to construct the scalar curvature, where we normalize it as in our previous study \cite{Wei2,Wei3}, while it is different from that of  Ref. \cite{Yang}.

For the $d$-dimensional Schwarzschild-AdS black hole, the corresponding normalized scalar curvature is
\begin{equation}
 R=-\frac{d-3}{2}\frac{4 \pi  T \hat{V}^{\frac{1}{d-1}}-d+3}{
   \left(2 \pi  T \hat{V}^{\frac{1}{d-1}}-d+3\right)^2},
   \label{scalar}
\end{equation}
with $\hat{V}=(d-1)V/\omega_{d-2} = r_h^{d-1}$. We see that $R$ has both  vanishing and divergent behaviours.
Since the small black hole branch is unstable (as shown above) we will not consider that branch. From (\ref{tempar}), we find that the black hole temperature is functions of the pressure $P$ and horizon radius $r_{h}$, so we can say that the normalized scalar curvature is dependent on the pressure $P$.

\begin{figure}
\includegraphics[width=7cm]{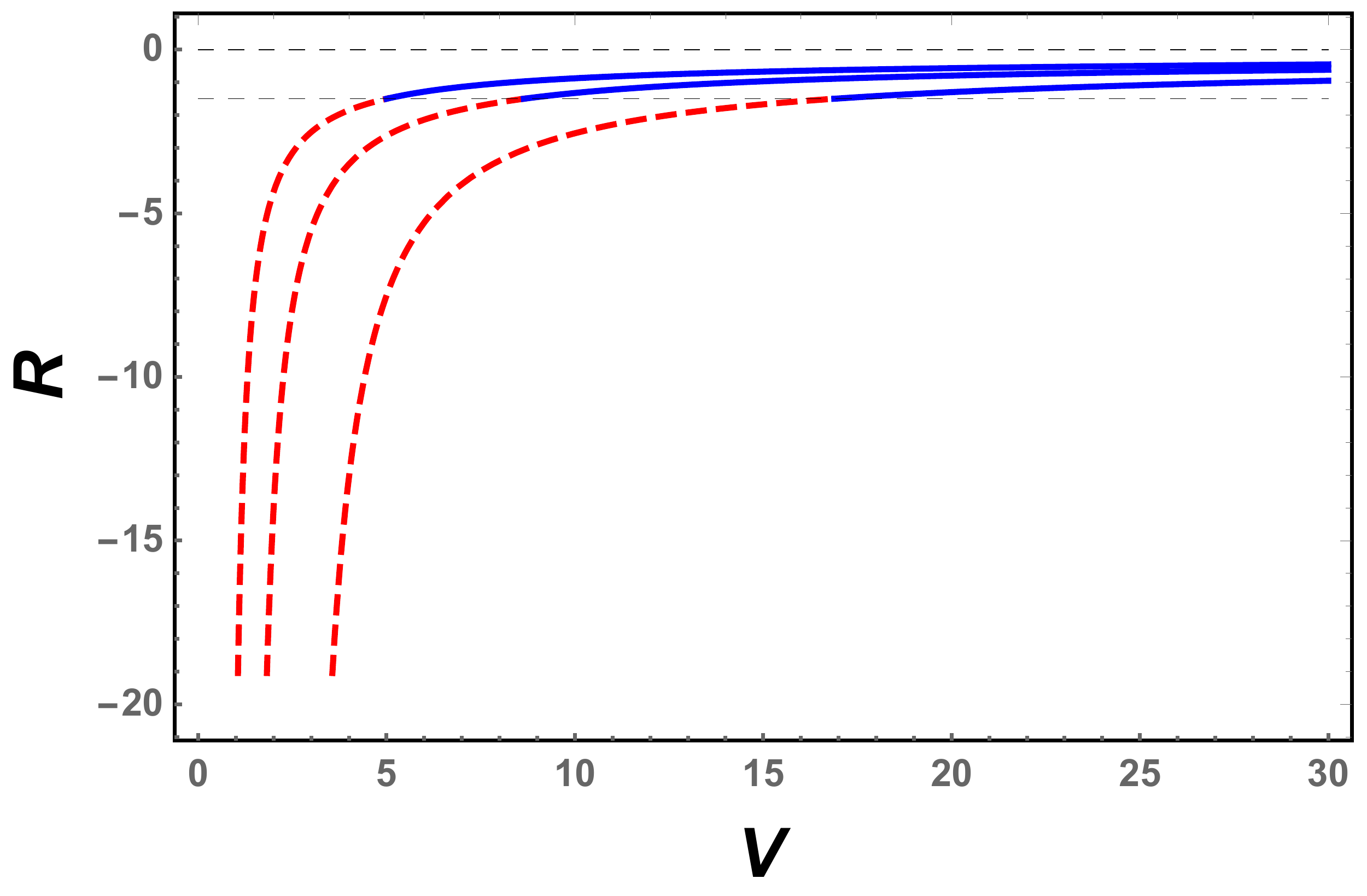}
\caption{The scalar curvature $R$ as a function of the volume $V$ for $d=4$ and $T$=0.2, 0.25, and 0.3 from bottom to top. The red dashed and blue solid curves are for the metastable and stable large black holes, respectively. Since the small black hole is unstable, we have excluded them in this picture.}\label{pd4}
\end{figure}

In Fig. \ref{pd4} we plot $R$ as a function of $V$ for $d=4$; in other dimensions  the behaviour is similar. The red dashed and blue solid curves correspond to metastable and stable large black holes respectively. For both of these  large black hole branches the scalar curvature is negative, and the attractive interaction always dominates for both. This result is not hard to understand:  in this spacetime there exists a competition between  thermal radiation and black hole phases. Thermal radiation has  effectively repulsive microstructure interactions and  so in order to form a black hole these must become attractive.  A detailed study also shows that at the black hole minimum temperature, the normalized scalar curvature diverges. If we conjecture that the absolute value of $R$ is an indicator of the intensity of the interaction, this black hole case is more like an ideal rigid body.

Note that if we exclude the metastable large black hole branch,  the divergent behavior is also removed. The curves then begin at the
HP phase transition points, and upon inserting the temperature (\ref{tpdh}) into the scalar curvature (\ref{scalar}), we obtain the amazing result
\begin{equation}
 R_{\text{HP}}=-\frac{(d-1)(d-3)}{2}, \label{constant}
\end{equation}
namely that the Ruppeiner curvature at the HP transition point is a universal constant, independent of the temperature, pressure, or the horizon radius, while only dependent on the dimension. We expect this constant may indicate  the fundamental nature of the HP phase transition, and we now consider one possible interpretation of this result.

In physics, there exist many thresholds, beyond or below which we have rather different physical behaviour. We contend that the HP transition is
one such threshold.  In  forming or destroying a massive black hole in the AdS space, the HP phase transition is the formation of a black hole from
pure thermal radiation.  At this point the repulsive microstructure interactions of the radiation must become attractive microstructure interactions for the
black hole.  The lack of dependence on thermodynamic parameters of the Ruppeiner curvature at this point strongly suggests a universal property in
the formation of black hole degrees of freedom.

{\it Summary}---We have shown that a $d$-dimensional Schwarzschild-AdS black hole exhibits a universal behaviour at the HP transition point.  For any given dimension, the system is characterized by two special temperatures: the HP phase transition temperature and the black hole minimum temperature. Although each is pressure dependent, we find an exact universal  relation (\ref{dual}) between them:  the minimum temperature $T_0$ in ($d$+1) dimensions equals to the HP phase transition temperature $T_{\text{HP}}$ in $d$ dimensions.  Similar to the AdS/CFT correspondence, the temperatures $T_0$ and $T_{\text{HP}}$ can be regarded as the physical quantities residing in the bulk and boundary respectively.  The duality between two HP phase transition temperatures in successive dimensions is realized by the metastable black hole branch, which has habitually been ignored. We expect that this dual relation discloses deeper fundamental aspects of the HP phase transition.

Furthermore, when applying methods of Ruppeiner thermodynamic geometry to  Schwarzschild-AdS black holes, we find attractive interactions dominate amongst the microstructures for both the metastable and stable large black hole branches. Moreover, the normalized scalar curvature at the HP phase transition point
is a universal dimension-dependent  negative constant. We suggest that this constant  can be understood as a pressure independent critical point
where a virtual black hole turns to a real one from the thermal radiation. Therefore, we presented a corresponding microscopic threshold in contrast to the macroscopic phase transition. This may be very helpful for understanding the black hole formation at the microscopic level, especially for supermassive black holes formed from the huge interstellar clouds of radiation as expected.

In summary, a novel dual relation between the black hole minimum temperature and HP phase transition temperature in two successive dimensions was discovered. Another universal constant indicating a threshold of the formation of a black hole in the microscopic level was also found. It would be interesting to apply the study of the HP phase transition to other black hole backgrounds. Of further interest is the study of the holographic aspects of the HP phase transition and universal underlying black hole microstructures with different thresholds.

Our results  for the dual relation can be extended to the charged case, where  the lowest temperature $T_0$ and the HP phase transition temperature are
\begin{eqnarray}
 T_0&=&\frac{2 \sqrt{(d-3) P \left(-2 (d-3) \Phi
   ^2+d-2\right)}}{\sqrt{\pi } (d-2)},\\
 T_{\text{HP}}&=&\frac{2 \sqrt{P \left(-2 (d-3) \Phi
   ^2+d-2\right)}}{\sqrt{\pi } \sqrt{d-1}},
\end{eqnarray}
in the grand canonical ensemble,  where the electric potential $\Phi$ is fixed. This implies
\begin{eqnarray}
 \frac{T_0}{T_{\text{HP}}}=\frac{\sqrt{(d-3) (d-1)}}{d-2},
\end{eqnarray}
which is independent of $\Phi$, and
 \begin{eqnarray}
{T_0}\left(d+1,\frac{\sqrt{(d-1)(d-3)}}{d-2}\Phi\right) =
 {T_{\text{HP}}}(d,\Phi)
\end{eqnarray}
is the generalization of (\ref{dual}) .

{\emph{Acknowledgements}.}---This work was supported by the National Natural Science Foundation of China (Grants No. 11675064, and No. 11875151)
and by the Natural Sciences and Engineering Research Council of Canada.

\end{document}